\documentclass[conference]{IEEEtran}
\IEEEoverridecommandlockouts
\usepackage{cite}
\usepackage{amsmath,amssymb,amsfonts}
\usepackage{algorithmic}
\usepackage{graphicx}
\usepackage{textcomp}
\usepackage{xcolor}
\usepackage[a4paper, total={184mm,239mm}]{geometry}

\usepackage{pifont}
\usepackage{subcaption}

\renewcommand{\IEEEbibitemsep}{0pt plus 0.5pt}
\makeatletter
\IEEEtriggercmd{\reset@font\normalfont\fontsize{7.35pt}{8.25pt}\selectfont}
\makeatother
\IEEEtriggeratref{1}

\def\BibTeX{{\rm B\kern-.05em{\sc i\kern-.025em b}\kern-.08em
    T\kern-.1667em\lower.7ex\hbox{E}\kern-.125emX}}

\begin{document}

\title{
DAISM: Digital Approximate In-SRAM\\Multiplier-based Accelerator for DNN\\Training and Inference
\thanks{This work was supported, in part, by JST CREST from Japan with Grant JPMJCR18K1.}
}

\author{
\IEEEauthorblockN{Lorenzo Sonnino\IEEEauthorrefmark{1},
Shaswot Shresthamali\IEEEauthorrefmark{1},
Yuan He\IEEEauthorrefmark{1},
Masaaki Kondo\IEEEauthorrefmark{1}${}^,$\IEEEauthorrefmark{2}}

\IEEEauthorblockA{\{lsonnino,~shaswot,~isaacyhe,~kondo\}@acsl.ics.keio.ac.jp}

\IEEEauthorblockA{\IEEEauthorrefmark{1}Keio University, Yokohama, Japan}
\IEEEauthorblockA{\IEEEauthorrefmark{2}RIKEN Center for Computational Science, Kobe, Japan}
}

\maketitle

\begin{abstract}
DNNs are widely used but face significant computational costs due to matrix multiplications, especially from data movement between the memory and processing units. One promising approach is therefore Processing-in-Memory as it greatly reduces this overhead. However, most PIM solutions rely either on novel memory technologies that have yet to mature or bit-serial computations that have significant performance overhead and scalability issues. Our work proposes an in-SRAM digital multiplier, that uses a conventional memory to perform bit-parallel computations, leveraging multiple wordlines activation. We then introduce DAISM, an architecture leveraging this multiplier, which achieves up to two orders of magnitude higher area efficiency compared to the SOTA counterparts, with competitive energy efficiency.

\end{abstract}



\section{Introduction} \label{sec:intro}
Deep Learning~(DL) has gained widespread popularity in recent years and become ubiquitous in many diverse applications, including daily tasks such as facial recognition, and applications that require extensive training like language models. As a result, many domain-specific accelerators have been proposed to streamline and optimize the computations for performance gains in terms of latency, energy, and chip area~\cite{eyeriss}. Typically, a large fraction of the computations for Deep Neural Networks~(DNNs) are general matrix multiplications~(GEMMs) and many accelerator designs have focused on accelerating them.

One method is to approximate the computations for performance gains by using approximate arithmetic~\cite{bitwise_or} or reduced precision~\cite{lower_part_or}. These methods leverage the inherent error resilience of Neural Networks~(NNs) to small computational errors. It arises primarily due to parameter over-provisioning and the independent distributed computations within each layer of the NN.

Another direction for optimizing GEMMs is to use Processing-In-Memory~(PIM). Since matrix multiplication is embarrassingly parallel, reading and transferring the data from memory to the processor consumes a lot of power and bottlenecks the entire computation pipeline~\cite{data_movement_1, data_movement_2}. PIM solutions perform computation directly in/near memory and thus minimize this bottleneck~\cite{racetrack, reram_analog}.

As attractive as PIM may be, current solutions have severe drawbacks that prevent their widespread adoption. For example, resistive memory-based designs are very sensitive to device-to-device variations and they also require conversion of data between analog and digital domains, which further drives up the energy cost and reduces throughput and accuracy. Furthermore, such analog computation-based technology requires significant changes in chip design and thereby incurs large design and manufacturing costs~\cite{why_not_analog_1, why_not_analog_3}.

Another alternative is in-memory bit-serial computation used, for example, by existing SRAM-based PIM technologies~\cite{sram_bit_serial_2, sram_bit_serial_3}. As a consequence, this requires the data to be reorganized in a bit-serial manner and incurs significant overhead in both performance and complexity. While latency issues from bit-serial operations may be alleviated through pipelining, the area overhead and complexity cannot be overlooked. Furthermore, since most of the existing computations are optimized for bit-parallel operation, bit-serial hardware is bound to lag behind in terms of efficiency while combining these two types introduces additional complexity in designing hardware. Finally, fundamental device- and circuit-level limitations, such as the current carrying capacity of a metal wire, also prevent bit-serial solutions to scale~\cite{serial_scale}.

In this work, we propose a novel in-SRAM approximate multiplier that brings the best of both worlds. Our multiplier performs matrix multiplications in memory thereby reducing the time and energy required for moving data. The multiplication is performed in a bit-parallel manner by using multiple wordlines activation that approximates multiplication with a simple bitwise OR, which is a perfect match to the tight and regular layout of conventional SRAM technology. The computational errors arising from this approximation are acceptable as DNNs are quite resilient due to over-provisioned parameters~\cite{acc_0}. It is possible to implement our multiplier in conventional SRAMs with minimal design modification and thus making it easily adopted in existing systems. We also propose DAISM - a DNN accelerator architecture that leverages our novel in-SRAM approximate multiplier. Our evaluations show that energy and performance gains can be obtained compared to other existing baselines.

The main contributions of the paper are:
\begin{itemize}
    \item We propose a novel in-SRAM approximate multiplier that approximates matrix multiplication with bitwise OR operation.
    \item We propose the DAISM architecture that leverages the in-SRAM approximate multiplier to realize performance gains and trade-offs between latency, area, and energy efficiency.
    \item We perform extensive evaluations on our proposed DAISM architecture and compare it with current SOTA baselines. Our results show that it is energy efficient, and requires fewer clock cycles with minimal to no degradation in model accuracy.
    \item We discuss and analyze the different trade-offs possible with our architecture.
\end{itemize}

This article is structured as follows. Section~\ref{sec:bckg} recap the basics of binary multiplication, then discusses some related work and how this work differs from previously published papers. Section~\ref{sec:mul} presents the basic concept behind the proposed approximate multiplier as well as some ways of improving its performances. Section~\ref{sec:acc} then introduces the DAISM architecture. Section~\ref{sec:eval} explains the evaluation methodology and discusses the multiplier and accelerator's performances. Finally, Section~\ref{sec:concl} concludes this work.

\section{Background and related works} \label{sec:bckg}
\subsection{Binary multiplication} \label{sec:bckg-bin}
To multiply two operands (the multiplicand and the multiplier), partial products (PP) first need to be generated. Each PP equals either the shifted multiplicand or $0$ if the corresponding bit from the multiplier is $0$. Those partial products then need to be added, which incurs significant overhead due to carry propagation.
Meanwhile, floating point (FP) numbers consist of 3 segments: sign, exponent, and mantissa, varying in size by data type. Multiplying them involves multiplying mantissas as unsigned integers and adding exponents. The mantissa is then normalized and the exponent is realigned. Finally, the output's sign bit is an XOR of the operands' sign bit.

\subsection{Related works} \label{sec:bckg-rw}
To enhance DL architecture performance, some new multipliers employ approximations~\cite{bitwise_or, lower_part_or} as DNNs have a large error resilience. For instance, \cite{bitwise_or} decreases PPs by performing bitwise OR operations among them. However, they still demand adder trees. \cite{lower_part_or} instead approximates the lower part of the result via PP bitwise OR, and the upper part using approximate Full-Adder logic. Still, none of these multipliers can operate in memory. Our approximate multiplier can, which solves the data movement problem.

Other works such as \cite{reram_analog, racetrack} use Processing-in-Memory (PIM), in which novel memory technologies are used for direct in-memory computations, bypassing processing units. Examples include \cite{racetrack} which employs RaceTrack memory for in-memory integer multiplication, and \cite{reram_analog} which leverages ReRAM memory for MAC operations. These technologies showcase minimal energy use, no data movement, and better performances by fully capitalizing on DNNs data parallelism. Nonetheless, these memory technologies face challenges to their novelty. RaceTrack and ReRAM, being yet-to-mature, lack the research and optimization of traditional SRAM. Analog in nature, they need digital-analog-digital conversions, impairing signal, increasing power use, and limiting throughput~\cite{why_not_analog_1, why_not_analog_3}. Our approach is based on conventional digital SRAM instead.

Finally, SRAM-based PIM technologies such as \cite{sram_bit_serial_2, sram_bit_serial_3} all perform bit-serial computations. They hence suffer from data reformulation and lack the support of SOTA hardware, as these are often bit-parallel. Bit-parallel multipliers instead benefit from the latest breakthrough and can be easily integrated into existing systems. Finally, in-memory bit-serial multiplication often has a very large complexity \cite{bit_serial_problem}. While this can be solved through pipelining, it comes with complexity and area overheads. Our architecture only uses bit-parallel hardware and can easily be implemented in existing technologies.

\section{Proposed multipliers} \label{sec:mul}
\subsection{Core concept} \label{sec:mul-core}
Carry propagation during partial summing decreases throughput and increases energy consumption. The proposed multiplier therefore avoids this by approximating this sum by a bitwise OR, requiring no adder tree, sacrificing some computational accuracy instead. Previous work proposed this as well, but only on the lower part of the result with no accuracy recovery mechanism~\cite{lower_part_or}. Furthermore, by using a slightly modified SRAM memory thoroughly described by \cite{sram_logic}, this step can be performed in memory. Indeed, by reading multiple wordlines at the same time, a bitwise OR between them is read instead. \cite{sram_logic} proved such technology to be viable and at a negligible cost as it only required some extra sense amplifiers. This SRAM can also function as a traditional memory but to allow multiple wordlines activation, a special address decoder must be designed though this will be proven to be negligible later on. Multiple wordlines activation has also been shown to pose no major problems in terms of signal-to-noise ratio or throughput~\cite{sram_logic}. Finally, the cost of the extra sense amplifier can be avoided by re-wiring the existing one in traditional SRAM.

\begin{figure}[bt]
    \vspace{-3.0mm}
    \centering
    \includegraphics[width=0.85\columnwidth]{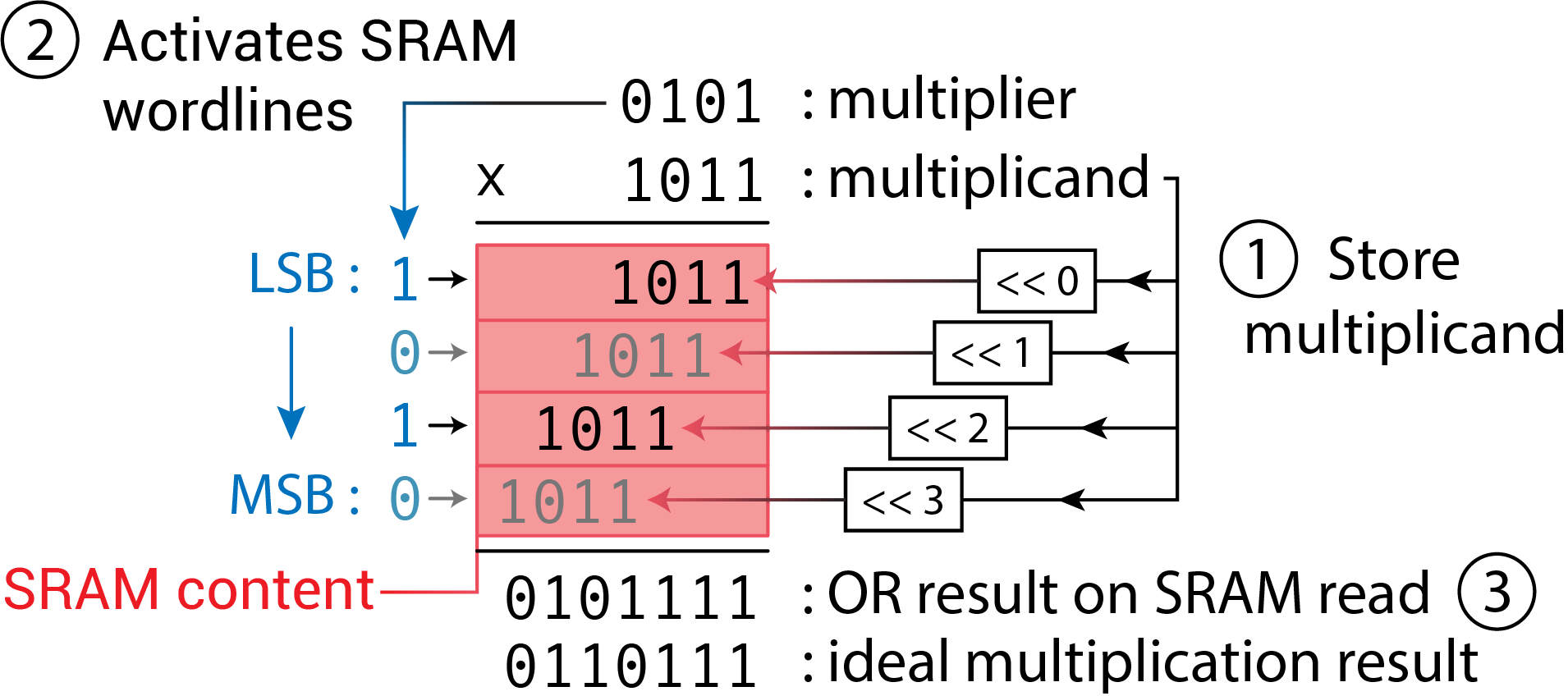}
    \caption{Example of the proposed multiplier's concept for $a = 1011$ and $b = 0101$. The SRAM line is read if the corresponding bit from the multiplier is 1}
    \label{fig:mul-concept}
\end{figure}

Fig. \ref{fig:mul-concept} describes the proposed multiplier. First, the multiplicand is stored. The multiplier is then used to activate multiple wordlines, generating PPs. By doing so, a wired-OR between PP is read, which approximates the result. This multiplier will be referenced as \texttt{FLA}, standing for \textit{Full Lines Activation}. For DNNs, the multiplicand is a kernel element, and the multiplier is an input element. The small size of most kernels makes a moderately-sized memory enough to store them, as will be discussed in Section \ref{sec:eval}. Finally, this approach makes handling data represented in two's complement difficult. This work however focuses on FP mantissa multiplication, which only uses unsigned integers.

\subsection{Storing pre-computed values} \label{sec:mul-ab}
\begin{figure}[bt]%
    \centering
    \includegraphics[width=0.65\columnwidth]{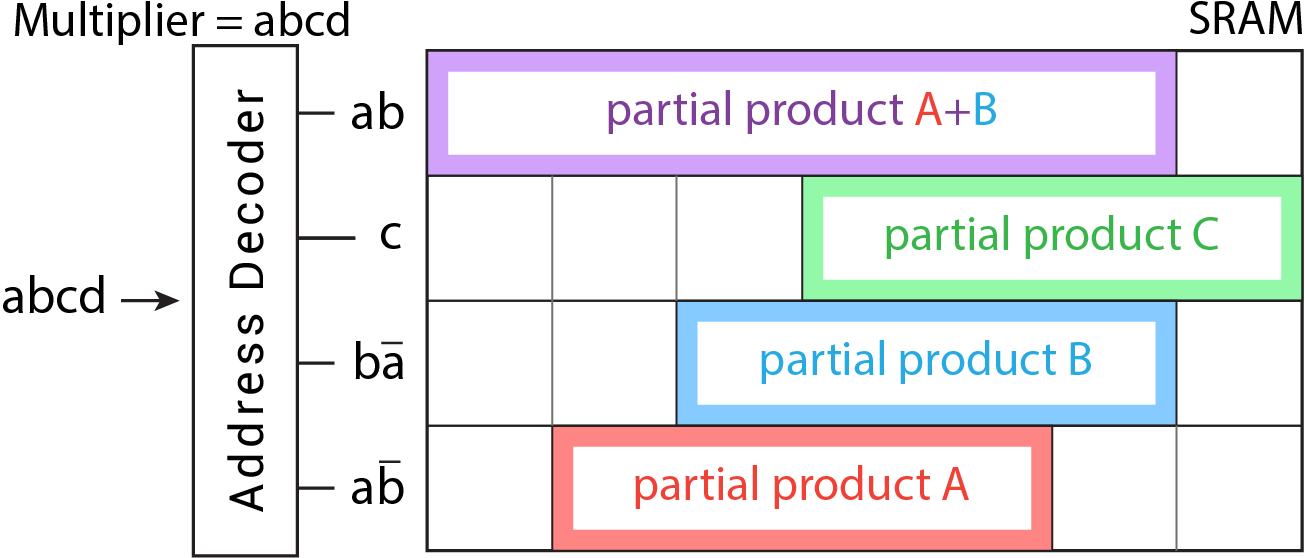}
    \caption{In \texttt{PC2}, the pre-computed sum between the two largest PP is stored}%
    \label{fig:mul-ab}%
\end{figure}

As previously stated, accuracy drops most when two successive lines must be activated. Let us assign capital letters to each PP. For instance, in an 8-bit setup, $A$ refers to the multiplicand shifted 7 times, and $H$ represents the unshifted multiplicand. Most of the accuracy loss happens when $A$ and $B$'s wordlines are active at the same time. Indeed, neighboring PPs have a high chance of collisions and those two directly affect the MSBs of the output. If instead of storing the LSB's PP ($H$ for 8-bit values) the exact result of $A+B$ is stored as shown in Fig. \ref{fig:mul-ab}, accuracy can be recovered with only a slightly more complex address decoder at no additional memory cost. This new PP would be selected whenever $A$ and $B$ should both be active. This is referenced as \texttt{PC2} standing for \textit{Pre-Computed} sums between 2 partial products. All other PPs are handled as in \texttt{FLA}. This article will explore \texttt{PC3} as well, in which the pre-computed sum of all possible combinations of the $A$, $B$, and $C$ lines are stored.

\subsection{Floating point generalization} \label{sec:mul-fp}
For FP arithmetic, the proposed multipliers are only capable of handling mantissa multiplication. The exponent and sign bits are handled separately, and multiplications by zero are bypassed.

Furthermore, the IEEE FP standard requires an extra ``$1$'' to be added at the MSB of the mantissa. This ``$1$'' is implicit in the binary representation. A 23-bit mantissa hence becomes a 24-bit unsigned integer whose MSB is $1$. The PP $A$ is hence active for all operands and, if PP $B$ must be activated as well, the $AB$ line (storing the pre-computed sum of these PPs) in \texttt{PC2} will be activated instead. The line for PP $B$ will hence never be active and can be left out, reducing memory consumption. \texttt{PC3} also greatly benefits from this, as many combinations between the $A$, $B$, and $C$ lines are no longer possible.

Moreover, because the proposed multiplier does not use any carry, the computation can be truncated arbitrarily, greatly improving performances at the cost of accuracy. This article will hence explore \texttt{PC2\_tr} and \texttt{PC3\_tr} in which the result is truncated to only compute the $n$ MSB. The value of $n$ is the mantissa width of the data type, including the leading ``$1$''.

Finally, many accelerators do not use standard FP but rather variations such as BFP. Because this multiplier handles arbitrary-size integer mantissa, any other FP representation can make use of this multiplier as long as it requires integer multiplications. This article explores the \texttt{float32} format as well as \texttt{bfloat16}~\cite{bfloat16}. The latter is similar to \texttt{float32} but uses a 7-bit mantissa instead of the standard 23. This number format is most notably used in Google's TPU.

\section{Accelerator architecture} \label{sec:acc}
\subsection{Core architecture} \label{sec:acc_arch}
\begin{figure}[bt]
    \vspace{-3.0mm}
    \centering
    \includegraphics[width=0.99\columnwidth]{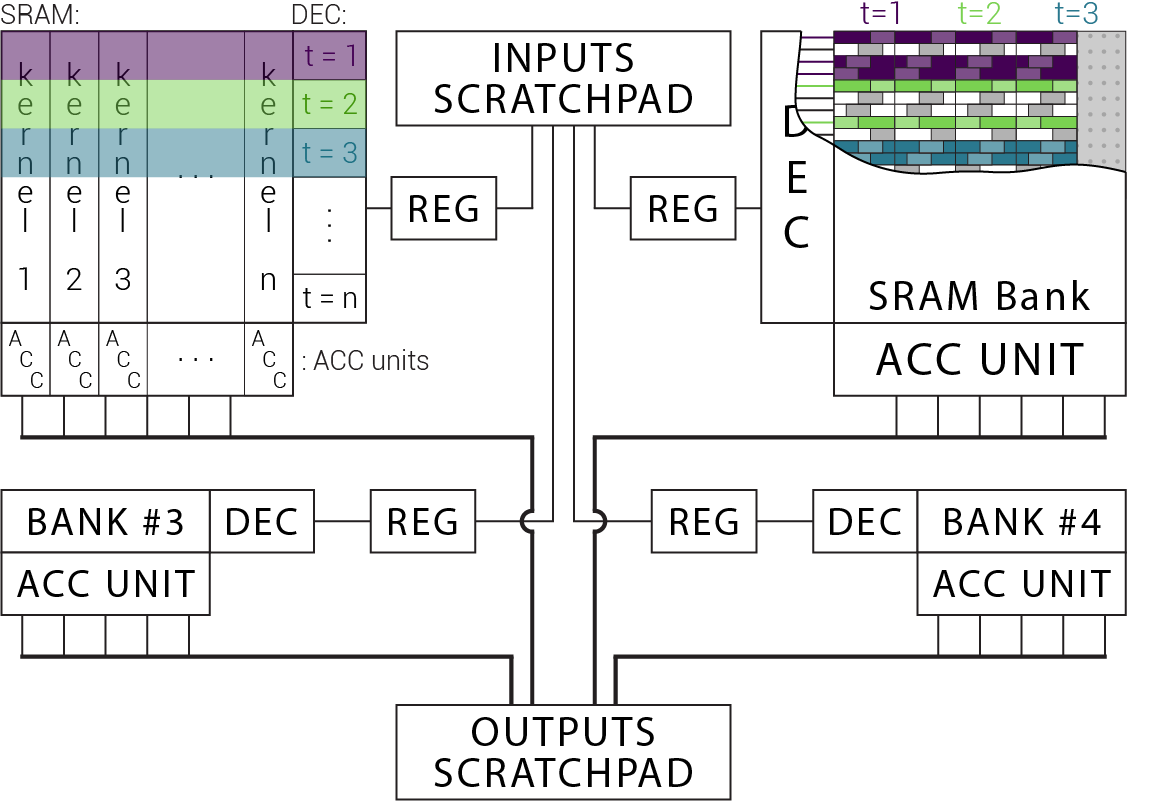}
    \caption{4 banks DAISM architecture. Inputs are fed one at a time to the SRAM from a register file through the address decoder. The dotted area represents unused SRAM space (not to scale)}
    \label{fig:arch}
\end{figure}

The proposed architecture (Fig.~\ref{fig:arch}) replaces the systolic array with a large SRAM memory, modified as proposed by~\cite{sram_logic} to support a wired-OR operation through multiple wordlines activation. Each kernel would be flattened and stored as shown in Fig.~\ref{fig:arch}. The inputs are taken from the top scratchpad and stored in a register file. They are then read one at a time and used to activate SRAM wordlines through an address decoder. Each input is hence multiplied by all the kernel elements on the same row at the same time. The results from these products are then fed to an accumulator at the bottom, accumulating the results of the multiplications. The final results are finally stored in another scratchpad memory.

This architecture can be applied to any variation of the proposed multiplier and any SRAM size.

\subsection{Architecture variations} \label{sec:acc_var}

A variation of this architecture involves dividing the large square SRAM memory into smaller square banks. This eases SRAM manufacturing and allows for different inputs to be fed to different banks simultaneously, as shown for the four banks in Fig.~\ref{fig:arch}.

The architecture also prefetches inputs from the scratchpad into an intermediary register file, like  \cite{eyeriss} does, except it only has one per bank. This reduces the frequency of expensive scratchpad reads, favoring the smaller register file.

This architecture is hence compatible and comparable to most systolic array architectures, only changing the way operands are multiplied and the way kernels are encoded.

For floating point numbers, this pipeline can only be used to multiply mantissa's as unsigned integers. The exponents must be handled separately, similar to how a block floating point architecture would work. This data type only has one exponent per matrix, reducing data size and improving performance.

\section{Evaluation} \label{sec:eval}
This section evaluates the multipliers shown in Table \ref{tab:eval-multipliers} in terms of energy consumption per computation, and accuracy loss.

\begin{table}[]
\small
\caption{Summary of the proposed multipliers}
\label{tab:eval-multipliers}
\centering
\begin{tabular}{|l|c|c|c|l|}
\hline
Config.          & Precomputed wordlines & Truncation \\ \hline\hline
\texttt{FLA}     & No                    & No         \\ \hline
\texttt{PC2}     & Between 2 PP          & No         \\ \hline
\texttt{PC3}     & Between 3 PP          & No         \\ \hline
\texttt{PC2\_tr} & Between 2 PP          & Yes        \\ \hline
\texttt{PC3\_tr} & Between 3 PP          & Yes        \\ \hline
\end{tabular}
\normalsize
\vspace{-2.5mm}
\end{table}

As a baseline multiplier for the energy consumption, the 32-bit floating point multiplier from \cite{fp32_mult} is assumed to be used in an architecture similar to Eyeriss \cite{eyeriss} to take into account operands read. \cite{fp32_mult} is chosen as a baseline multiplier as it provides energy consumption and area for different levels of truncation. The proposed multipliers are evaluated for both \texttt{float32} and \texttt{bfloat16} operands. The multiplier from \cite{fp32_mult} must hence be adapted to approximate the energy consumption of a \texttt{bfloat16} multiplier, as will be explained in Section \ref{sec:eval-en-meth}.

The proposed architecture is evaluated in terms of on-chip area and performance compared to Eyeriss, using Accelergy, as will be explained in Section \ref{sec:eval-acc-meth}.


\subsection{Accuracy} \label{sec:eval-acc}

\subsubsection{Methodology} \label{sec:eval-acc-meth}
The accuracy drop is evaluated on large models trained on ImageNet, such as ResNet-50~\cite{resnet, imagenet}. The baseline uses \texttt{float32} data while the proposed architectures use \texttt{bfloat16}.

\subsubsection{Results} \label{sec:eval-acc-res}
\begin{figure}
    \centering
    \includegraphics[width=0.75\columnwidth]{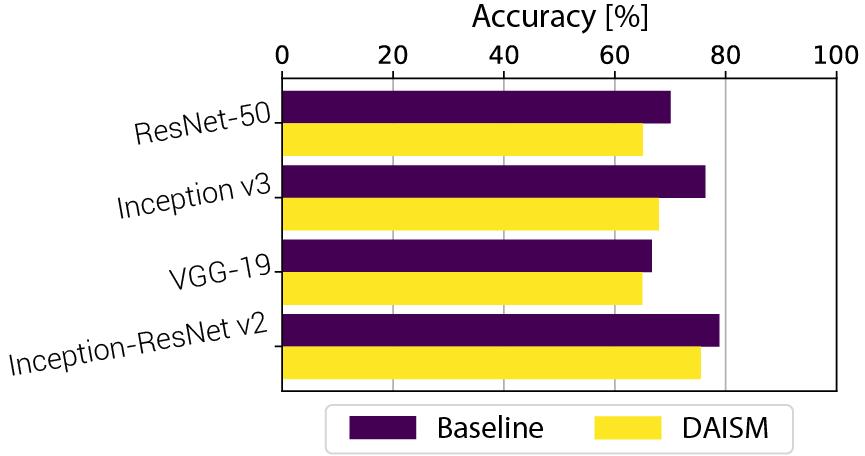}
    \caption{Accuracy evaluation for larger CNN using \texttt{bfloat16} truncated \texttt{PC3} compared to an exact \texttt{float32} baseline}
    \label{fig:eval-acc-larger}
\end{figure}

Fig. \ref{fig:eval-acc-larger} shows the accuracy for various large CNNs when executed on \texttt{PC3} compared to the FP32 baseline. Despite some accuracy drop being felt compared to the \texttt{float32} baseline, DAISM is still able to achieve high accuracy on larger models while bringing energy and performance benefits as will be discussed in the following sections.

Finally, DNN inference with approximate computing is especially targeted toward edge devices that rarely employ deeper neural networks. The choice of accelerating floating point mantissa arithmetic also limits error magnitude (as opposed to integer arithmetic or exponent handling) while still providing great benefits as will be shown in Sections~\ref{sec:eval-en} and \ref{sec:eval-arch}.


\subsection{Energy consumption} \label{sec:eval-en}

\subsubsection{Methodology} \label{sec:eval-en-meth}
The energy consumption of the proposed multipliers has been evaluated with CACTI \cite{cacti_1, cacti_2} and Synopsys's Design Compiler using NANGATE 45nm technology. Our multipliers are compared to the truncated \texttt{float32} multiplier from \cite{fp32_mult}. For both the proposed and the baseline multipliers, operands read has been considered.

Finally, a baseline \texttt{bfloat16} multiplier can be deduced by scaling \cite{fp32_mult} as in \eqref{eq:bf16_mul} in which the energy consumptions $E_\text{sim,16}$ and $E_\text{sim,32}$ have been simulated using NANGATE 45nm (\texttt{bfloat16} and \texttt{float32} respectively) and $E_\text{16}$, $E_\text{32}$ are the energy consumptions of the baseline multipliers (\texttt{bfloat16} and \texttt{float32} respectively).

\begin{equation}
\begin{split}
    E_\text{16} &= E_\text{32} \cdot \frac{E_\text{sim,16}}{E_\text{sim,32}} \cdot T
    \label{eq:bf16_mul}
\end{split}
\end{equation}

\subsubsection{Results} \label{sec:eval-en-res}
\begin{figure}[htb]
    \centering
    \includegraphics[width=0.90\columnwidth]{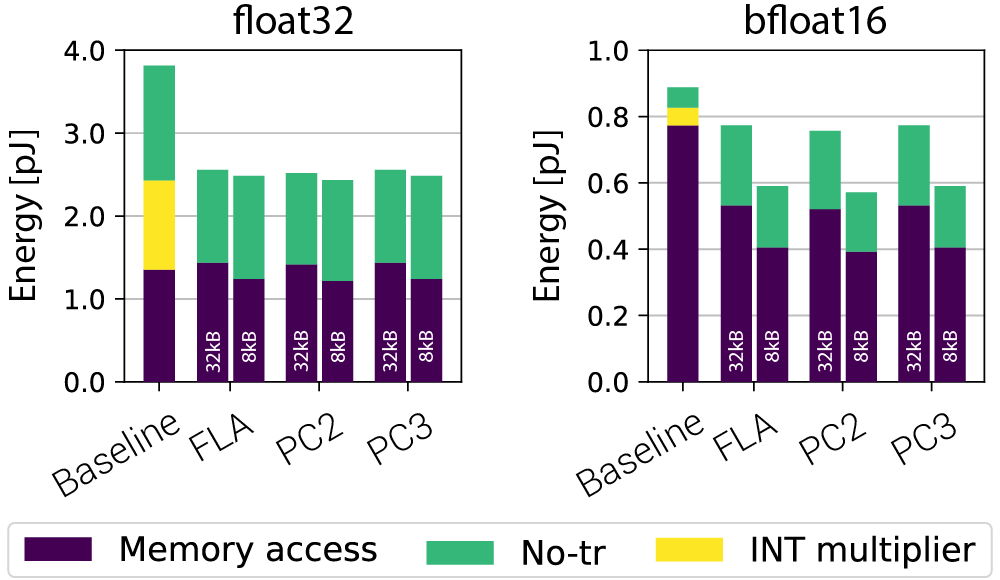}
    \caption{Energy break-down for all the proposed mantissa multipliers compared to a common baseline for either a 32kB or an 8kB SRAM. \textit{No-tr} represent the spared energy consumption by truncation}
    \label{fig:energy_breakdown}
\end{figure}

Fig. \ref{fig:energy_breakdown} shows the energy consumption for all the proposed multipliers compared to the baseline. This includes the energy consumption required for additional components, such as the address decoder. The figure compares the energy consumption per computation across proposed multipliers, datatype, and the size of the bank performing in-memory computations. We can see the following points from Fig. \ref{fig:energy_breakdown}:
\begin{enumerate}
    \item The cost of the address decoder is negligible. It represents less than $0.5\%$ of the energy consumption in all cases.
    \item Memory read plays an important role in energy consumption. For in-memory computations, this cost is reduced by the many reuses and by reading one operand from a register file.
    \item While using a smaller memory decreases energy consumption per read, the decrease in the number of computations per memory read cancels out the benefits. There is hence no major difference in terms of energy consumption per computation.
    \item Truncation allows for drastically improved performances as it nearly doubles the number of computations per memory read. Truncation therefore nearly halves memory read energy consumption, greatly reducing overall energy consumption.
\end{enumerate}

The In-Memory multipliers require fewer data movements to perform a multiplication operation. Indeed, one operand can be left in place, unlike a traditional multiplier. While this has been shown to have a significant impact on energy consumption\cite{data_movement_1, data_movement_2}, it has not been taken into account as when integrated into a DL accelerator, data movement is still required between the SRAM and an accumulator.

\begin{figure}[tb]
    \centering
    \includegraphics[width=0.85\columnwidth]{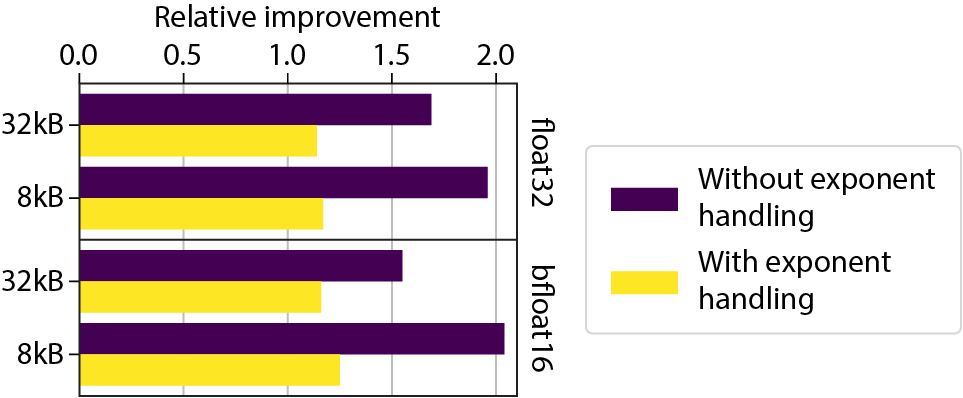}
    \caption{Relative improvement in energy consumption when taking into account exponent handling for different SRAM bank sizes and data types}
    \label{fig:energy_rel_imp}
\end{figure}

Exponent adding and realignment are common costs for both the baseline and the proposed multipliers. Adding this common cost reduces the benefits realized by using the proposed multipliers. Fig. \ref{fig:energy_rel_imp} shows the improvement in energy consumption when using \texttt{PC3\_tr} compared to the baseline.

Finally, the cost of pre-loading data is made negligible by the large operands reuse. For instance, The first layer of VGG-8 has 150,528 inputs for 1728 kernel elements. This means that each input is reused for a very large number of kernel elements and each kernel element is reused for thousands of inputs, making the cost of any pre-loading negligible.


\subsection{Architecture evaluation} \label{sec:eval-arch}

\subsubsection{Methodology} \label{sec:eval-arch-meth}
The proposed architecture and its variations (with a varying number of banks and memory size) are compared to the Eyeriss architecture \cite{eyeriss} using Accelergy and Timeloop \cite{accelergy}. VGG-8 will be used to evaluate the architecture as it is widely used and allows us to better highlight the key differences between the architectures as it uses a larger number of processing elements and require a larger amount of memory.

All architectures use the \texttt{bfloat16} datatype. The area of a truncated \texttt{bfloat16} has been computed the same way the energy consumption has been computed in \ref{sec:eval-en-meth}.

Finally, in-memory technologies have not been evaluated as they often perform full integer MAC operations and are therefore unable to perform floating point operations.

\subsubsection{Results} \label{sec:eval-arch-res}
\begin{figure}[tb]
    \centering
    \includegraphics[width=0.65\columnwidth]{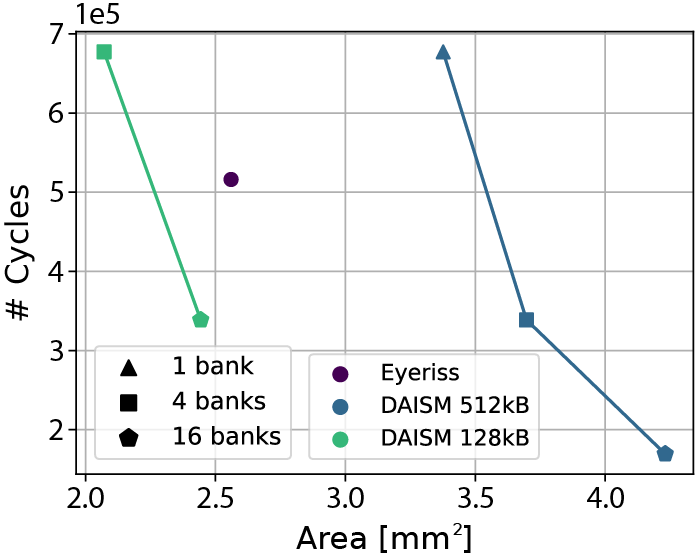}
    \caption{Architectures performances comparison when executing the first layer of VGG-8 in \texttt{bfloat16} representation between the proposed \texttt{PC3\_tr}-based architecture with different 45nm variations}
    \label{fig:arch-perf}
\end{figure}

Fig. \ref{fig:arch-perf} shows the trade-off between the number of cycles and the on-chip area to execute the first layer of VGG-8 on different architectures. The proposed architecture has been evaluated by using one single 512kB or 8kB SRAM memory, then by splitting it into smaller square banks.

\begin{figure}[tb]
    \centering
    \includegraphics[width=0.65\columnwidth]{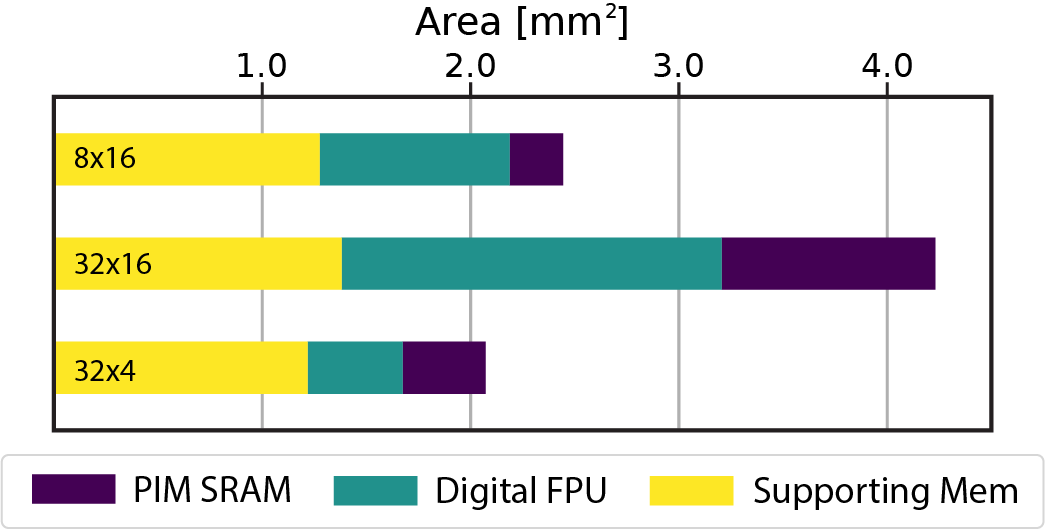}
    \caption{Detailed area breakdown of the DAISM architecture}
    \label{fig:eval-arch-area}
\end{figure}

Single bank architectures suffer from the lack of inputs that can be fed at each cycle. Indeed, some input elements must not be multiplied by all kernel elements, which decreases utilization. Moreover, while DAISM suits any memory shape, a standard squared memory is assumed. While such a 512kB bank can store up to 128x256 kernel elements, the considered layer only has 1728, leaving most of the memory unused.

Furthermore, the 1x512kB architecture can only use 128 kernel elements at a time. Hence, dividing the SRAM memory into smaller square banks, each taking distinct inputs at each cycle, decreases the number of cycles at the expense of some on-chip area and a larger data bus connecting the scratchpad to the SRAM banks increasing costs. As a consequence, the 16-bank design has 512 processing elements which are about 3x those of Eyeriss. This however requires more hardware for exponent handling and accumulation.

Decreasing the total on-chip memory allows an increase in the number of banks while maintaining a small on-chip area. This makes the 16 banks of 8kB variation the smallest architecture while maintaining the same performance as the 128kB bank one.

In Fig.~\ref{fig:eval-arch-area}, the main SRAM's area relative to other required digital circuits (exponent handling, accumulators) for each processing element is shown. When the SRAM's width is increased, its area is squares quadratically while the number of PE increases linearly. The opposite is true when the number of banks increases instead. Therefore, as memory banks get larger, the area becomes dominated by the SRAM memory with little performance benefits. However as the number of banks increases, the area becomes dominated by other digital circuits, and a larger cost is associated with each bank.

Table~\ref{tab:arch-comp} compares DAISM to Z-PIM~\cite{sram_bit_serial_2} and T-PIM~\cite{sram_bit_serial_3} which both use digital in-SRAM computation logics. Despite using 45nm technology, DAISM achieves comparable energy efficiency but up to two orders of magnitude higher overall performance and area efficiency with 16-bit inputs and weights. On the other hand, this advantage in computation density over Z-PIM and T-PIM remains an order of magnitude higher even if the operating frequency of DAISM is scaled down to 200MHz.

\begin{table}[bt]
\caption{Performances comparison between different PIM architectures}
\label{tab:arch-comp}
\centering\footnotesize


\begin{tabular}{|l|cc|c|c|}
\hline
Architecture      & \multicolumn{2}{c|}{DAISM}              & Z-PIM~\cite{sram_bit_serial_2} & T-PIM~\cite{sram_bit_serial_3} \\ \hline\hline
Config            & \multicolumn{1}{c|}{16x8kB} & {16x32kB} & \textemdash                    & \textemdash                    \\ \hline
Computations      & \multicolumn{2}{c|}{bit-parallel}       & bit-serial                     & bit-serial                     \\ \hline
Node {[}nm{]}     & \multicolumn{2}{c|}{45}                 & 65                             & 28                             \\ \hline
Area {[}mm$^2${]} & \multicolumn{1}{c|}{2.44} & {4.23}      & 7.57                           & 5.04                           \\ \hline
GE Area$^{\S}$ {[}mm$^2${]} & \multicolumn{1}{c|}{3.81} & {6.61}      & 5.91                           & 15.51$\sim$24.83                           \\ \hline
Clock {[}MHz{]}   & \multicolumn{2}{c|}{1000}               & 200                            & 50$\sim$280                    \\ \hline
Supply {[}V{]}    & \multicolumn{2}{c|}{1.0}                & 1.0                            & 0.75$\sim$1.05                 \\ \hline
GOPS              & \multicolumn{1}{c|}{502.52} & {1005.04} & 1.52$\sim$16.0$^{\ast}$        & 5.56$^{\dag}$                  \\ \hline
GOPS/mW           & \multicolumn{2}{c|}{0.23}               & 0.31$\sim$3.07$^{\ast}$        & 0.13$\sim$1.26$^{\ddag}$       \\ \hline
GOPS/mm$^2$       & \multicolumn{1}{c|}{205.68} & {237.55}  & 0.53$\sim$5.31$^{\ast}$        & 1.1$^{\dag}$                   \\ \hline
\end{tabular}

\vspace{1ex}
{\raggedright \hspace{2ex} $^{\S}$~Gate Equivalent area computed using nodes from \cite{gate_equivalent} \par}
{\raggedright \hspace{2ex} $^{\ast}$~Varies according to the weight sparsity~(0.1$\sim$0.9). \par}
{\raggedright \hspace{2ex} $^{\dag}$~Measured with input sparsity of 0.9 and weight sparsity of 0.5. \par}
{\raggedright \hspace{2ex} $^{\ddag}$~Varies with the input sparsity~(0.1$\sim$0.9); weight sparsity is set to 0.5. \par}

\normalsize
\vspace{-2.5mm}
\end{table}


\subsection{Final analysis} \label{sec:eval-recap}
Prior sections outlined trade-offs in the proposed multipliers and their impact.

First, between the evaluated multipliers, \texttt{PC3} is the best choice for three reasons:
\begin{enumerate}
    \item \texttt{PC3} has better accuracy;
    \item \texttt{PC3} requires fewer simultaneously active wordlines;
    \item The cost in terms of energy consumption per computation is similar.
\end{enumerate}

Truncation minimally affects accuracy, but significantly enhances energy efficiency per computation due to increased computations per memory read. Additionally, our bit-parallel approximate multiplier employs multiple wordlines for storing partial products, temporarily using more SRAM. This doesn't pose a major problem as SRAM is abundant and can accommodate many kernels at runtime (each kernel may take around a few tens of bytes). Moreover, when batch size is large during inference, it amortizes the cost of populating SRAM with the shifted bit patterns.

Finally, The proposed architecture improves performance through more processing elements and reduces energy consumption compared to Eyeriss due to lower per-computation energy. A trade-off exists between performance and on-chip area, which can be fine-tuned by selecting an appropriate number of banks and memory size. Table \ref{tab:summary} summarises the key benefits of the proposed multiplier and accelerator compared to other technologies.
\begin{table}[h]
\caption{Summary of the key differences between the DAISM accelerator and related work}
\label{tab:summary}
\centering\footnotesize
\begin{tabular}{|l|l|l|l|l|l|}
\hline
 &
  \textbf{\begin{tabular}[c]{@{}l@{}}Data \\ \scriptsize{Movement} \end{tabular}} &
  \textbf{\begin{tabular}[c]{@{}l@{}}Type of \\ \scriptsize{Computation} \end{tabular}} &
  \textbf{\begin{tabular}[c]{@{}l@{}}Memory \\ \scriptsize{Technology} \end{tabular}} &
  \textbf{\begin{tabular}[c]{@{}l@{}}Memory \\ \scriptsize{Reads} \end{tabular}}\\ \hline\hline
\textit{\textbf{DAISM}}                                                 & \textbf{None} & \textbf{Digital} & \textbf{Legacy} & \textbf{Single} \\ \hline
\textit{\begin{tabular}[c]{@{}l@{}}Digital \\ Multipliers\end{tabular}} & Required      & Digital          & Legacy          & Single \\ \hline
\textit{\begin{tabular}[c]{@{}l@{}}Analog \\ PIM\end{tabular}}          & None          & Analog           & Novel           & Single \\ \hline
\textit{\begin{tabular}[c]{@{}l@{}}SRAM \\ Digital PIM\end{tabular}}    & None          & Digital          & Legacy          & Multiple \\ \hline
\end{tabular}\normalsize
\vspace{-2.5mm}
\end{table}

\section{Conclusion} \label{sec:concl}
In this article, we propose multiple variations of an approximate digital in-SRAM multiplier for multiple data types. Most notably, the \texttt{PC3\_tr} variation stores pre-computed values, then uses an in-memory wired-OR to combine them, approximating the result. This allows for a decrease in energy consumption compared to a traditional multiplier while avoiding large accuracy drops. Finally, an accelerator architecture has also been introduced, capitalizing on this multiplier. Through our comprehensive evaluations, this accelerator has been shown to outperform Eyeriss, a cutting-edge accelerator, for a comparable chip area, and it is more area efficient than two SOTA SRAM-based PIM counterparts.


\bibliographystyle{IEEEtran}
\bibliography{main}

\end{document}